\documentstyle[12pt]{article}
\date{}
\begin{document}
\newcommand{\ba}{\begin{array}}
\newcommand{\ea}{\end{array}}
\newcommand{\sta}{\stackrel}
\newcommand{\ovl}{\overline}
\newcommand{\eee}{\mbox{e}}
\newcommand{\triex}{\mbox{\hspace{3ex}}}
\newcommand{\lla}{\longleftarrow}
\newcommand{\lra}{\longrightarrow}
\newcommand{\edoc}{\end{document}}
\newcommand{\pal}{\partial}
\newcommand{\Pfaff}{\mbox{Pfaff}\;}
\newcommand{\Pf}{\mbox{Pf}\;}
%
\thispagestyle{empty}

COND-MAT/9711156\\[5ex]
\mbox{}

\begin{center}
{\LARGE  Grassmann Variables and Exact Solutions for\\*[1.2ex]
Two-Dimensional Dimer Models}
\end{center}
\begin{center}
{\large R.\ Hayn \\[0.8ex]
\em MPA Elektronensysteme, Technische Universit\"at Dresden,\\
    D-01062 Dresden, Germany  \\[1.2ex]
\em and \\[1ex]  V.\ N.\ Plechko\\[0.8ex]
\em Bogoliubov Laboratory of Theoretical Physics \\[3pt]
    Joint Institute for Nuclear Research \\[3pt]
    141980 Dubna, Moscow Region, Russia\\}
\end{center}
\vfill
\noindent{\large A talk presented at the VIII International Conference
on Symmetry Methods in Physics, JINR, Dubna, Russia, July 28 -
August 2, 1997, to appear in the proceedings}\\*[8ex]
$\dagger$  E-mail: roland@tmps06.mpg.tu-dresden.de\\
$\ddagger$ E-mail: plechko@thsun1.jinr.dubna.su

\thispagestyle{empty}
%
\newpage
\setcounter{page}{1}
%
%
%
\title{\LARGE Grassmann Variables and Exact Solutions for Two-Dimensional
Dimer Models}

\author{
R. Hayn$\,^\dagger$ and V. N. Plechko$\,^{\ddagger}$ \\
\em  $\dagger$ MPA Elektronensysteme, Technische Universit\"at Dresden, \\
\em  D-01062 Dresden, Germany  \\[2mm]
\em  $\ddagger$ Bogoliubov  Laboratory of Theoretical Physics, \\
\em  Joint Institute for Nuclear Research, \\
\em  141980 Dubna, Moscow Region, Russia \\ }

\maketitle

\begin{abstract} We discuss some aspects of a new noncombinatorial
fermionic approach to the two-dimensional dimer problem in statistical
mechanics based on the integration over anticommuting Grassmann variables
and factorization ideas for dimer density matrix. The dimer partition
function can be expressed as a Gaussian fermionic integral. For regular
lattices, the analytic solution then follows by passing to the momentum
space for fermions.  \end{abstract}

%
%
\section{Introduction}

The dimer problem was originally formulated to describe the entropy of
diatomic molecules adsorbed on a crystal surface [1]. In a more abstract
sense, the two-dimensional (2D) dimer model is one of only few nontrivial
exactly solvable models in statistical mechanics [2-8]. In its original
state, this is a purely combinatorial problem, close in its inherent spirit
to the 2D Ising model [4]. The exact solution for the closed-packed 2D
dimer problem on the standard rectangular lattice was first obtained by a
direct combinatorial method [2,3]. A number of more complicated dimer
lattices have also been analyzed [5,6]. Traditional combinatorial
approaches of this kind, however, are somewhat complicated, and are
different, in their spirit, from the methods commonly used in condensed
matter physics. In this report we comment on a simple fermionic approach to
the 2D dimer problem based on the integration over anticommuting Grassmann
variables (nonquantum fermionic fields) and the mirror-ordered
factorization procedure for the dimer density matrix [7,8]. Neither
transfer-matrices nor combinatorics are used. The factorization method
rather resembles the idea of insertion of the Dirac unity,
$\Sigma\,|a\left> \right<a|=1$, when making transformations in quantum
mechanics. We first reformulate the combinatorial dimer partition function,
$Q$, in terms of the commuting nil-potent $(\eta)$-variables, and then
introduce into the resulting expression the anticommuting $(a)$-variables
by factorization of the local bond Boltzmann weights. Eliminating the
$\eta$-variables in the resulting $(\eta,a)$-representation, we come to a
purely fermionic expression for the same partition function $Q$.  Even for
the inhomogeneous distribution of the dimer weights over the lattice, the
partition function is expressible as a Gaussian fermionic integral.
Equivalently, the closed-packed 2D dimer problem is a theory of free
fermions on a lattice. For regular lattices, the analytic solution then
follows by transformation to the momentum space for fermions.

%
%
%
\section{Grassmann variables}

Let us remember first the basic rules of anticommuting analysis [9].
Grassmann variables are classic (nonquantum) fermionic numbers purely
anticommuting to zero. Given a set of Grassmann variables $a_{1},...,\,
a_{N}$, we have: $a_{i}a_{j}+ a_{j}a_{i} =0$, $a_{j}^{\,2}=0$. The first
important identity is that for the product of linearly transformed
variables:
$$
b_1\,b_2\,....\,\,b_N\,=\,\det A \cdot a_1\,a_2\,...\,a_N\,, \triex
b_i\,=\,\sum\limits_{j=1}^{N}\,A_{ij}^{}\,a_j\,,
\eqno(2.1)
$$
where $A$ is the matrix of transformation $a \to b$. The determinant
here appears due to the known interrelations between fermionic algebra
and determinant combinatorics.  The rules of fermionic integration were
first formulated by Berezin [9].  For one variable, we have:
$$
\int\,da_{j}^{}\cdot a_{j}= 1\,,\triex\,\int\,da_{j}^{}\cdot 1 = 0\,.
\eqno(2.2)
$$
In the multiple fermionic integral the differential symbols $da_{1},...,
\,da_{N}$ are again anticommuting with each other and with the variables
[9]. Due to the property $a_{j}^{2}=0$, any natural function defined on
the set $a_1,\ldots,a_N$ can be written as a finite polynomial in these
variables:
$$
f\,(a_1,a_2,...,a_N)=f_0\,+\!\!\!
\sum\limits_{(1\le j \le N)}\!\! f_j\,a_j\,+\,
\cdots\,+ f_{123...N}\;a_1a_2...a_N\;,
\eqno(2.3)
$$
where $f_0,\ldots,f_{123...N}$ are numerical parameters. Integrating
a polynomial function like (2.3) according to (2.2), we obtain:
$$
\int da_N\ldots da_2da_1\,f\,(a_1,a_2,\ldots,a_N) = f_{\,123\ldots N}\,,
\eqno(2.4)
$$
where $f_{\,123...N}$ is just the coefficient in the last term of the
polynomial (2.3). We see that fermionic integration is a trivial problem
if the integrand function is already known in a polynomial form. However,
this may be not the case in applications. From (2.1) and (2.4), one can
deduce the rules of change of variables in a fermionic integral under a
linear substitution.  As compared with the rules of a proper commuting
analysis the only difference is that the Jacobian will now appear in the
inverse power.

An important role in applications play Gaussian fermionic integrals
[9,10].  The Gaussian integral of the first kind is related to the
determinant:
$$
\int\prod\limits_{j=1}^{N}\,da_{j}^{\,*}da_{j}\,\exp\left(
\sum\limits_{i=1}^{N}\sum\limits_{j=1}^{N}a_{i}A_{ij}a_{j}^{\,*}
\right)\,=\,\det\hat{A}\;,
\eqno(2.5)
$$
where $\{a_{j},\,a_{j}^{*}\}$ is a set of purely anticommuting Grassmann
variables, the matrix $\hat{A}$ is arbitrary. By convention, the variables
$a_{j}^{}$ and $a_{j}^{\,*}$ can be considered as complex-conjugated
fermionic fields, otherwise these are independent variables. The fermionic
exponential is assumed in the sense of its series expansion, the series
terminates at some stage due to the property $a_{j}^{\,2}=0$. The
correspondent finite polynomial also can be obtained by multiplying
elementary factors $\exp\,(a_{i}A_{ij} a_{j}^{*})= 1+a_{i}A_{ij}a_{j}^{*}$.
The Gaussian integral of the second kind, for real fermionic fields, is
related to the Pfaffian:
$$
\int da_N\,...\,da_2da_1\,\exp\left(
\frac{1}{2}\,\sum\limits_{i=1}^{N}\sum\limits_{j=1}^{N}
a_{i}A_{ij}a_{j}\right)\,=\,\mbox{Pfaff}\,\hat{A}\,,
\hspace{3ex} A+A^{T}=0\,,
\eqno(2.6)
$$
where the matrix $\hat{A}$ is now skew-symmetric ($A_{ij}+A_{ji}=0$,
$\,A_{ii}=0\,$).  The Pfaffian is some combinatorial polynomial in elements
$A_{ij}^{}$ well known in mathematics for a long time. In fact, the
Pfaffian combinatorics is identical with that of fermionic Wick's theorem.
The Pfaffian and determinant of the associated skew-symmetric matrix are
algebraically related: $\det\,\hat{A}= (\,\mbox{Pfaff}\,\hat{A} \,)^{\,2}$.
This identity can be most readily proved just in terms of integrals
like (2.5) and (2.6), making use of a suitable change of variables in
(2.5).

%
%
\section{The dimer problem}

Let us consider the closed-packed 2D dimer model on a rectangular lattice
net with the most general inhomogeneous distribution of dimer weights
[7,8]. The dimers are objects living on lattice bonds. A given bond may
be either free or covered by a dimer together with the two adjacent lattice
sites. The closed-packing condition means that each lattice site must be
occupied by one and only one dimer. The lattice must have an even number
of sites to be covered completely by dimers in a closed-packed fashion. An
example of a closed-packed dimer configuration on a rectangular net can be
seen in Fig.~1a in [7].

Let the lattice sites be numbered by the integer Cartesian coordinates
$mn$, where $m=1,\ldots,M$, $n=1,\ldots,N$ run in horizontal and vertical
directions, respectively.  ${M}{N}$ is the total number of sites. We
define $t_{mn}^{(1)}$ and $t_{mn}^{(2)}$ to be the dimer weights
(activities) for horizontal $(mn\,|\, m{+}1n)$ and vertical $(mn\,|\,
mn{+}1)$ bonds, respectively. The weight of a free bond is $1$. The dimer
partition function is:
$$
Q=\sum\limits_{\{N\}}\,\Big\{\prod\limits_{m=1}^{M}
\prod\limits_{n=1}^{N}\,{t_{mn}^{(1)}}^{N_{mn}^{(1)}}\,
{t_{mn}^{(2)}}^{N_{mn}^{(2)}}\,\Big\}\,,
\eqno(3.1)
$$
where $N_{mn}^{(1)}$ and $N_{mn}^{(2)}$ are dimer occupation numbers
taking the values $1,0$, depending of whether a given bond is covered
or free, and the sum is taken over all closed-packed configurations.
Notice that in a completely homogeneous and symmetric case, $t_{mn}^{(1)}=
t_{mn}^{(2)}=1$, the partition function yields simply the total number of
closed-packed dimer configurations for a given lattice. The thermodynamic
interpretation is only possible if there are few sorts of dimers with
different weights (energies).

We need to transform $Q$ into a fermionic Gaussian integral. The
combinatorial definition (3.1) is not very suitable in this respect since
the symbol of the sum in (3.1) is only a conventional prescription for
counting over closed-packed configurations.  In fact, we need some kind of
a formal averaging which will select the closed-packed configurations
automatically. This can be realized in terms of the commuting nil-potent
variables [7]. With each lattice site we now associate the commuting
nilpotent variable $\eta_{mn}$, such that $\eta_{mn}^{\,2}=0$, and
write:
$$
Q = \!\!\!\ba[t]{c}Sp\vspace{-0.5ex} \cr (\eta) \ea
\Big\{\prod\limits\limits_{m=1}^{M}\prod\limits\limits_{n=1}^{N}
(1\!+\!t_{mn}^{(1)}\eta_{mn}\eta_{m+1\,n})
(1\!+\!t_{mn}^{(2)}\eta_{mn}\eta_{mn+1})\,\Big\}\;,
\eqno(3.2)
$$
with the averaging rules for one variable:
$$
\ba[t]{c} \mbox{Sp} \vspace{-1ex} \cr
\mbox{$\scriptstyle{(\eta_{mn})}$} \ea
\,(\,1\,|\,\eta_{mn}\, |\,\eta^{2}_{mn}\,|\,
\eta^{3}_{mn}\,|\,...\,)=\,
(\,0\,|\,1\,|\,0\,|\,0\,|\,...)\,.
\eqno(3.3)
$$
The global $\eta$-averaging in (3.2) assumes local averagings like (3.3)
at all sites. In (3.2) we also assume the free-boundary conditions:
$\eta_{M+1\,n}= \eta_{m\,N+1}=0$. Expression (3.2) exactly reproduces the
combinatorial partition function (3.1). The product of the Boltzmann
factors $1+t\eta\eta'$ forming the density matrix in (3.2) generates all
possible dimer coverings of a lattice by `dimer molecules',
$\,\eta\,t\,\eta'$, with no restrictions, while the $\eta$-averaging
according to (3.3) selects the closed-packed configurations. These are the
configurations in which each lattice site is covered by one and only one
`atom' or `monomer', $\eta_{mn}\,$, which can only be a part of some dimer.

%
%
\section{Fermionization}

Let it be given two Grassmann variables, $a$ and $a^{*}$. The elementary
Gaussian exponential is $\eee^{\,\lambda aa^{*}}=1+ \lambda aa^{*}$, where
$\lambda$ is a parameter. Noting (2.2), for the associated Gaussian
averages we then find: $\int da^{*}da\, \eee^{\lambda aa^{*}}\{
1,a,a^{*},aa^{*}\}$ $=\{\lambda,\,0,\, 0,\,1\,\}$. Making use of these
rules, we can factorize the weight factors from (3.2) as is shown in (4.1)
below. For the whole lattice, we introduce a set of purely anticommuting
Grassmann variables, $\{\,a_{mn}^{}, a_{mn}^{*}, b_{mn}^{}, b_{mn}^{*}\}$,
and write:
$$
\ba{llr}
1+t_{mn}^{(1)}\eta_{mn}\eta_{m+1n}
\\[1ex]
\,=\,\int\limits_{}^{}da_{mn}^{*}da_{mn}\,\mbox{e}^
{\mbox{$a_{mn}a_{mn}^{*}$}}
(1+t_{mn}^{(1)}\,a_{mn}^{}\eta_{mn}^{})\,
(1+a_{mn}^{*}\eta_{m+1n}^{})
\\[1ex]
=\!\! \ba[t]{c}Sp \vspace{-0.5ex} \\ (a_{mn}) \ea \!\!
\{\,A_{mn}^{}\,A_{m+1n}^{*}\,\}\,,
\cr\ea
\eqno(4.1a)
$$  $$
\ba{llr}
1+t_{mn}^{(2)}\eta_{mn}\eta_{mn+1}
\\[1ex]
\,=\,\int\limits_{}^{}db_{mn}^{*}\,db_{mn}^{}\,\mbox{e}^
{\mbox{$b_{mn}b_{mn}^{*}$}}
(1+t_{mn}^{(2)}\,b_{mn}^{}\eta_{mn}^{})\,(1+b_{mn}^{*}
\eta_{mn+1}^{})\,
\\[1ex]
= \!\!  \ba[t]{c}Sp \vspace{-0.5ex} \\ (b_{mn}) \ea \!\!
\{\,B_{mn}^{}\,B_{mn+1}^{*}\,\}\,.
\cr\ea
\eqno(4.1b)
$$
In the last lines we introduce the abbreviated notation for the arising
factors (to be called shortly Grassmann factors), $A_{mn}, B_{mn},
A_{m+1n}^{*}, B_{mn+1}^{\,*}$, while $Sp\,(...)$ stands for the symbol of
the Gaussian fermionic averaging. Neglecting the averaging symbol, the
Boltzmann weights from (3.2) are now presented in a factorized form:
$A_{mn}^{} A_{m+1n}^{*}\,$, $\,B_{mn}^{} B_{mn+1} ^{\,*}$. In general, for
the whole lattice, there are four Grassmann factors, $A_{mn}, A_{mn}^{*},
B_{mn}, B_{mn}^{\,*}$, which all involve the same variable ${\eta}_{mn}$
associated with a given $mn$ site. The idea of the next step is to place
nearby the above four factors and to average over $\eta_{mn}$ in each
group independently thus passing to a purely fermionic expression for $Q$.
The obstacle to this method is that the individual Grassmann factors are
neither commuting nor anticommuting with each other. It might, therefore,
be difficult, in general, to find the four relevant factors nearby in a
global product. The problem of a suitable ordering of the non-commuting
Grassmann factors thus arises. This ordering problem can be solved applying
the mirror-ordered factorized procedure [7]. The related ideas were first
developed in the context of the 2D Ising model [11,12]. The result is that
the density matrix from (3.2) can be represented in the following
`mirror-ordered' factorized form [7,8]:
$$
Q\,(\eta) = \ba[t]{c}Sp\vspace{-0.5ex} \cr (a,b) \ea
\Big\{\,\prod\limits_{n=1}^{N}\,\Big[\,\prod\limits_{m=1}^{M}\,
\stackrel{\sta{m}{\longleftarrow\!-}}{B_{m\,n}^{\,*}}\,\cdot\,
\prod\limits_{m=1}^{M}\,
\stackrel{\sta{m}{-\!-\!-\!-\!\longrightarrow}}
{A^{\,*}_{mn}B_{mn}}\!A_{mn}\,\Big]\,\Big\}\,.
\eqno(4.2)
$$
The $\eta$-averaging can be performed at the junction of the $m$-ordered
products in (4.2) [7]. This results in a purely fermionic expression for
the partition function, which can in turn be simplified by integrating
out the extra fermionic variables. In this way we obtain [7]:
$$
Q=\int\prod\limits_{n=1}^{N}\prod\limits_{m=1}^{M}
\sta{\sta{m}{\longleftarrow}}{dc_{mn}}
\mbox{exp}\,\Big\{\,\sum\limits_{m=1}^{M}\sum\limits_{n=1}^{N}
\left[\,t_{mn}^{(1)}c_{mn}c_{m+1n} +
(-1)^{m+1}t_{mn}^{(2)}c_{mn}c_{mn+1}\,\right]\Big\}\;,
\eqno(4.3)
$$
with $c_{M+1\,n}=c_{m\,N+1}=0$, where $c_{mn}$ are new totally
anticommuting Grassmann variables. The product in the measure is ordered
as follows. First, we multiply $dc_{mn}$ over $m$, with fixed $n$, then we
multiply the resulting products over $n$. In the field-theoretical
language, the fermionic form in the exponential is called action, since the
action is quadratic in fermions, we deal here with a free-fermion field
theory on a lattice. It is also possible to eliminate the sign factor
$(-1)^{m+1}$ from the action. This can be done by the rescaling of the
variables in the integral: $c_{mn} \to c_{mn}\, i^{\,m^2+3/2}$,  $\,dc_{mn}
\to dc_{mn}\,i^{\,-m^2-3/2}$. This yields equivalent representation [8]:
$$
Q=\int\prod\limits_{n=1}^{N}\prod\limits_{m=1}^{M}
\sta{\sta{m}{\longleftarrow}}{dc_{mn}}
\mbox{exp}\,\Big\{\,\sum\limits_{m=1}^{M}\sum\limits_{n=1}^{N}
\Big[\,t_{mn}^{(1)}c_{mn}c_{m+1n} +
\,i\,t_{mn}^{(2)}c_{mn}c_{mn+1}\,\Big]\Big\}\;.
\eqno(4.4)
$$
The dimer partition function is now expressed as a simple fermionic
Gaussian integral. The exact solution for a regular dimer lattice can now
be obtained  by passing to the momentum space for fermions (Fourier
substitution). This is illustrated in the next section with examples of
regular (homogeneous) rectangular and brick-hexagonal dimer lattices. It
is interesting that these two models exhibit quite different properties.
While there is no phase transition for the standard rectangular dimer
lattice, there is an exotic phase transition with frozen low-temperature
phase for the hexagonal lattice.

%
%
\section{Analytic results for regular lattices}

Let us consider the 2D dimer model on the standard homogeneous rectangular
lattice. Assuming $t_{mn}^{(1)}= t_1$, $t_{mn}^{(2)} =t_{2}^{}$ in (4.4),
the partition function is:
$$
Q\, ={\displaystyle\int}\,
\prod\limits_{n=1}^{N}\prod\limits_{m=1}^{M}\,
\sta{m}{\sta{\longleftarrow}{dc_{mn}^{}}}\, \exp\,
\sum\limits_{m=1}^{M}\sum\limits_{n=1}^{N}\,
\left[\,t_{1}\,c_{mn}^{}c_{m+1\,n}+\,i\,t_{2}\,c_{mn}^{}c_{m\,n+1}^{}\,
\right]\,,
\eqno(5.1)
$$
where $c_{M+1n}=0, c_{mN+1}=0$. We deal here with a finite lattice with
a free boundary of size $M\times N$, with ${M}{N}$ sites. In what follows
we assume that $M$ and $N$ both are even. It is interesting that integral
(5.1)  can be performed exactly despite of that the translational
invariance is broken by free boundary. The diagonalizing Fourier
substitution is [7,8]:
$$
c_{mn}\,=\,\frac{ 2\,i^{\;m+n+\frac{3}{2}}}{\sqrt{(M+1)(N+1)}}\,
\sum\limits_{p=1}^{M}\sum\limits_{q=1}^{N}\,
c_{pq}\,\sin\left(\frac{\pi pm}{M+1}\right)\,
\sin\left(\frac{\pi qn}{N+1}\right)\,.
\eqno(5.2)
$$
In the momentum space, we find [8]:
$$
Q= \int \prod\limits_{q=1}^{N}\prod\limits_{p=1}^{M}
\sta{\sta{p}{-\!\rightarrow}}{dc_{pq}}\,
\exp\,\sum\limits_{p=1}^{M}\sum\limits_{q=1}^{N}\,
\Big[\,c_{\bar{p}\bar{q}}\,c_{pq}\,\Big(\,t_{1}\cos\frac{\pi p}{M+1}+
it_{2}\cos\frac{\pi q}{N+1}\Big)\Big]\,,
\eqno(5.3)
$$
where $c_{\bar{p}\bar{q}}$, $c_{pq}$ are the new variables of integration,
$\bar{p}=M{+}1{-}p$, $\bar{q}=N{+}1{-}q$ are `conjugated' momenta selected
by the orthogonality relations for the Fourier eigenfunctions from (5.2).
Notice that the variables $c_{\bar{p}\bar{q}}$, $c_{pq}$ enter,
simultaneously, into the $pq$ and $\bar{p}\bar{q}$ terms of the total sum
in (5.3) (in the given case, these terms appear to be equal to each other).
To single out explicitly the true independent variables (which is needed to
perform the integral) we have to combine these $pq$ and $\bar{p}\bar{q}$
terms together and then to reduce the sum to a half-interval with respect
to $M$ or $N$ (choice between $M$ or $N$ is here arbitrary) in order to
avoid the double counting of the same terms. Making the reduction with
respect to $M$, the partition function becomes:
$$
Q= \int \prod\limits_{p=1}^{\frac{1}{2}M}\prod\limits_{q=1}^{N}
dc_{pq}dc_{\bar{p}\bar{q}}\,\exp\,\sum\limits_{p=1}^{\frac{1}{2}M}
\sum\limits_{q=1}^{N}\Big[c_{\bar{p}\bar{q}}\,c_{pq}\,
\Big(\,2t_{1}\cos\frac{\pi p}{M+1}+ 2it_{2}\cos\frac{\pi q}{N+1}\Big)
\Big]\,.
\eqno(5.4)
$$
The value of the integral readily follows from elementary rules like
(2.2).  In the resulting product we then make the $q \leftrightarrow
\bar{q}$ (or $i \leftrightarrow -i$) symmetrization, and find:
$$
Q=\prod\limits_{p=1}^{\frac{1}{2}M}\prod\limits_{q=1}^{\frac{1}{2}N}
\,\left[\,4t^{2}_{1}\cos^2\frac{\pi p}{M+1}+
4t^2_{2}\cos^2\frac{\pi q}{N+1}\,\right]\,, \triex
\mbox{ even } M\,,N\,.
\eqno(5.5)
$$
This is the exact solution for the 2D dimer model on a finite rectangular
lattice with free boundary. This result was first obtained by the
combinatorial method [2,3].

Taking the limit of infinite lattice, we obtain the free energy per site
for the standard rectangular dimer lattice:
$$
-\beta f_{\rm REC}^{}
= \frac{1}{M\!N}\ln Q\;\Big|_{\,M\! N\to \infty}^{}
=\frac{1}{\pi^2}\int\limits_{0}^{\frac{\pi}{2}}\!
\int\limits_{0}^{\frac{\pi}{2}}dp\,dq\,\ln
\left[\,4\,t_{1}^{2}\cos_{}^{2}p+4\,t_{2}^{2}\cos_{}^{2}q\,\right]\,.
\eqno(5.6)
$$
The free energy (5.6) can be represented in other equivalent forms, in
particular, it can be written in the form [8]:
$$
-\,\beta\,f_{\rm REC}^{}=\,\frac{1}{4}\ln\,(t_1t_2) +
\frac{G}{\pi} +\frac{1}{2\pi}\,
\int\limits_{t_2}^{\,t_1}\!\int\limits_{t_2}^{\,t_1}\,
\frac{dx\,dy}{x^2+y^2}\,,
\eqno(5.7)
$$
where $G=0.915\,965\,594$ is Catalan's constant. The term $G/\pi$
corresponds to a purely combinatorial contribution to the free energy (and
entropy) while the thermodynamics is governed by the doubled-integral term.
In thermodynamic interpretation, the dimer weights are to be chosen in the
form $t_\alpha=\exp\,(\varepsilon_\alpha)$, $\varepsilon _\alpha= \beta
E_\alpha$, $\alpha=1,2$, where $E_\alpha$ are the dimer energies (more
precisely, the true energies are $-E_\alpha$), and $\beta=1/kT$, where
$kT$ is the temperature in energy units. A nontrivial thermodynamic
interpretation is only possible for the nonequal dimer energies (weights),
$E_1\neq E_2$ ($t_1\neq t_1$). If $t_1=t_2$, the energy of any
closed-packed dimer configuration is the same and there is no
thermodynamics. The thermodynamic functions can all be deduced from the
free energy. There is no phase transition in the dimer model on a
rectangular lattice [2,3,6,8]. In particular, the specific heat is a smooth
function of temperature getting its maximum at $(kT/\Delta E)_{\rm\,max}=
0.477572$, at which point $(C/k)_{\rm\,max}= 0.169375$ [8], where $\Delta E
=|E_1-E_2|$, $C/k$ is the dimensionless specific heat, $k$ is the Boltzmann
constant.

Let us now consider the closed-packed dimer model on the standard hexagonal
lattice. There are three different weights, $t_1,t_2,t_3$, connected to
each site. The hexagonal lattice is equivalent to the so-called brick
lattice which can be obtained from a rectangular lattice net by
the elimination of some of the vertical bonds [6,7]. Therefore, the
fermionic integral for $Q$ for the hexagonal lattice can be deduced from
(4.3) and/or (4.4).  Another possibility, which we follow below, is to
interpret the hexagonal lattice (yet on a rectangular net) in a diagonally
layered fashion, as it was done for the 2D Ising model in [12].  Within
such interpretation, omitting further details, we obtain the fermionic
integral for the partition function of the 2D hexagonal dimer model in a
particularly simple form [8]:
$$
Q\,=\int\prod\limits_{mn}^{} dc_{mn}^{*}dc_{mn}^{}\,\exp\,
\sum\limits_{mn}^{}\,\left[ c_{mn}^{}\left(t_1\,c_{mn}^{\,*} -
t_2\,c_{m-1n}^{\,*}- t_3\,c_{mn-1}^{\,*}\right)\right]\,,
\eqno(5.8)
$$
where $c_{mn}, c_{mn}^{\,*}$ are anticommuting Grassmann variables.
Assuming the periodic closing conditions for fermions, which in the given
case is the boundary approximation, we then pass to the momentum space by
the simplest Fourier substitution:
$$
c_{mn}\,=\,\frac{1}{\sqrt{L^2}}\,\sum\limits_{pq}^{}\,
c_{pq}^{}\,\eee^{\,i\frac{2\pi}{L}(mp+nq)}\,, \;\;\;
c_{mn}^{\,*}\,=\,\frac{1}{\sqrt{L^2}}\,\sum\limits_{pq}\,c_{pq}^{\,*}\,
\eee^{\,-i\frac{2\pi}{L}(mp+nq)}\,.
\eqno(5.9)
$$
The integral (5.8) becomes:
$$
Q= \int\prod\limits_{pq}^{} dc_{pq}^{\,*}dc_{pq}^{}\,\exp\,
\sum\limits_{pq}^{}\left[\,c_{pq}^{}c_{pq}^{\,*}
\left( t_1 - t_2\,\eee^{\,i\,\frac{2\pi}{L}p} -
t_2\,\eee^{\,i\,\frac{2\pi}{L}q}\right)\right]\,,
\eqno(5.10)
$$
with evident explicit solution for $Q$. Respectively, the free energy per
site for the hexagonal lattice appears in the form:
$$
-\beta\,f_{\rm HEX}^{}
= \frac{1}{2}\,\int\limits_{0}^{2\pi}\int\limits_{0}^{2\pi}
\frac{dp\,dq}{(2\pi)^2}\,\ln\,\left[\;t_1 -t_2\,\eee^{\,i\,p}
-t_2\,\eee^{\,i\,q}\,\right]\,.
\eqno(5.11)
$$
In the thermodynamic interpretation, the weights are: $t_\alpha=\exp\,
(\varepsilon_\alpha)$, where $\varepsilon_ \alpha=\beta E_\alpha$ are the
dimensionless energies, the true energies of dimers are $-E_\alpha$, and
$\beta=1/kT$. A nontrivial thermodynamics is only possible if the dimer
energies (weights) are not all equal to each other.  From (5.11) it follows
that there is the phase transition, with frozen low-temperature phase, in
the dimer model on the hexagonal lattice. To be definite, let $t_1 \leq t_2
< t_3$ (or $-E_3 < -E_2 \leq -E_1$), the critical temperature is then given
by the condition $t_1+t_2 -t_3=0$. At low temperatures, $t_1+t_2<t_3$, the
system is frozen. In this phase ($T<T_c$) all dimers are captured at the
bonds with the highest weight (the lowest energy) and the specific heat
identically equals zero. In the high-temperature phase, $t_1+t_2>t_3$, the
system is unfrozen. In this phase ($T>T_c$) the specific heat exhibits the
$|T-T_c|^{-\frac{1} {2}}$ singularity as $T$ approaches $T_c$ from above.
In the $T>T_c$ phase the specific heat is [8]:
$$
\left.\frac{C}{k}\;\right|_{\rm\,HEX} = \frac{1}{\pi}\,\frac{
A\,(\varepsilon_1,\varepsilon_2,\varepsilon_3)}{ \mathstrut \sqrt{
(t_1+t_2+t_3)(-t_1+t_2+t_3)(t_1-t_2+t_3)(t_1+t_2-t_3) }}\,,
\eqno(5.12)
$$
where the numerator is a nonsingular function of temperature:
$
A\,(\varepsilon_1,\varepsilon_2,\varepsilon_3) =
(\varepsilon_1-\varepsilon_2) (\varepsilon_1-\varepsilon_3)\,t_{1}^{2}+
(\varepsilon_2-\varepsilon_1) (\varepsilon_2-\varepsilon_3)\,t_{2}^{2}+
(\varepsilon_3-\varepsilon_1) (\varepsilon_3-\varepsilon_2)\,t_{3}^{2}\,.
$
The unusual phase transition in the hexagonal dimer model with frozen
$T<T_c$ phase was first mentioned by Kasteleyn in 1963, for a recent
comprehensive discussion see [6]. The anisotropic specific heat in the
form (5.12) is cited here from [8].

%
%
\section{Conclusions}

We have discussed some aspects of the fermionic interpretation of the 2D
closed-packed dimer problem. In its original state, the 2D dimer model is
rather a discrete combinatorial problem. Introducing the fermionic
variables into the original partition function, we add, effectively, some
amount of new degrees of freedom to the original discrete sum. In a sense,
this extension of the inherent symmetry just makes the dimer problem
tractable analytically. The dimer partition function can be transformed
into a fermionic Gaussian integral even for the inhomogeneous distribution
of dimer activities over the lattice bonds. For simple homogeneous
lattices, the analytic solutions then follow by passing to the momentum
space for fermions.



\begin{thebibliography}{33}

\bibitem{gug52}{}          
E.A. Guggenheim, Mixtures (Clarendon Press, Oxford, 1952) Chapter~X.

\bibitem{kast61}{}         
P.W. Kasteleyn, Physica (Utrecht) {\bf 27} (1961) 1209.

\bibitem{fish61}{}          
M.E. Fisher, Phys. Rev. {\bf 124} (1961) 1664.

\bibitem{mont64}{}          
E.W. Montroll, \ In: Applied Combinatorial Mathematics,
Ed. by E.F. Beckenbach (New York, John Willey, 1964).

\bibitem{sam80}{}          
S. Samuel, J. Math. Phys. {\bf 21} (1980) 2820.

\bibitem{nagle90}{}         
J.F. Nagle, C.S.O.~Yokoi and S.M. Bhattacharjee, \ In: Phase Transitions
and Critical Phenomena, Eds. C. Domb and J.L. Lebowitz, Vol.13 (London,
Academic Press, 1990).

\bibitem{hayn93}{}          
R. Hayn and V.N. Plechko, J. Phys. A  {\bf 27} (1994) 4753.

\bibitem{hayn94}{}          
R. Hayn and V.N. Plechko, in preparation.

\bibitem{ber66}{}           
F.A. Berezin, The Method of Second Quantization (New York, Academic,
1966).

\bibitem{ber69}{}           
F.A. Berezin, Uspekhi Matem. Nauk, {\bf 24} (3) (1969) 3.

\bibitem{ple85}{}           
V.N. Plechko, Sov. Phys. Doklady {\bf 30} (1985) 271.

\bibitem{ple88}{}            
V.N. Plechko, Physica A {\bf 152} (1988) 51.

\end{thebibliography}
\end{document}